\documentclass[letter]{article}
\usepackage{moriond_letter2,epsfig}
\usepackage[OT1]{fontenc}
\usepackage{multitoc} 
\usepackage{wrapfig}
\usepackage{color}
\usepackage{amssymb}

\def\be{\begin{equation}}
\def\ee{\end{equation}}
\def\bea{\begin{eqnarray}}
\def\eea{\end{eqnarray}}

\def\clover{C$_{\ell}$over}
\def\deg{$^{\circ}$}

\begin{document}

\title{Detecting the B-mode Polarisation of the CMB with \clover}

\author{C.~E.~North$^1$, B.~R.~Johnson$^1$, P.~A.~R.~Ade$^2$,
  M.~D.~Audley$^3$, C.~Baines$^4$, R.~A.~Battye$^4$, M.~L.~Brown$^3$,
  P.~Cabella$^5$, P.~G.~Calisse$^2$, A.~D.~Challinor$^{6,7}$,
  W.~D.~Duncan$^8$, P.~G.~Ferreira$^1$, W.~K.~Gear$^2$, D.~Glowacka$^3$,
  D.~J.~Goldie$^3$, P.~K.~Grimes$^2$, M.~Halpern$^9$, V.~Haynes$^4$,
  G.~C.~Hilton$^8$, K.~D.~Irwin$^8$, M.~E.~Jones$^1$,
  A.~N.~Lasenby$^3$, P.~J.~Leahy$^4$, J.~Leech$^1$, B.~Maffei$^4$,
  P.~Mauskopf$^2$, S.~J.~Melhuish$^4$, D.~O’Dea$^3$,
  S.~M.~Parsley$^2$, L.~Piccirillo$^4$, G.~Pisano$^4$,
  C.~D.~Reintsema$^8$, G.~Savini$^2$, R.~Sudiwala$^2$, D.~Sutton$^1$,
  A.~C.~Taylor$^1$, G.~Teleberg$^2$, D.~Titterington$^3$,
  V.~Tsaneva$^3$, C.~Tucker$^2$, R.~Watson$^4$, S.~Withington$^3$,
  G.~Yassin$^1$, J.~Zhang$^2$ \\}

\address{\vspace{0.5cm}
  $^1$\footnotesize Astrophysics, University of Oxford, Oxford, UK \\
  $^2$\footnotesize School of Physics and Astronomy, Cardiff University, UK\\
  $^3$\footnotesize Cavendish Laboratory, University of Cambridge, Cambridge, UK\\
  $^4$\footnotesize School of Physics and Astronomy, University of Manchester, UK\\
  $^5$\footnotesize Dipartimento di Fisica, Universit\`a di Roma Tor Vergata, Italy\\
  $^6$\footnotesize Institute of Astronomy, University of Cambridge, UK\\
  $^7$\footnotesize DAMTP, University of Cambridge, UK\\
  $^8$\footnotesize National Institute of Standards and Technology, USA\\
  $^9$\footnotesize University of British Columbia, Canada\\
}

\maketitle

\abstracts{ We describe the objectives, design and predicted
  performance of \clover, which is a ground-based experiment to
  measure the faint \mbox{``B-mode''} polarisation pattern in the
  cosmic microwave background (CMB).  To achieve this goal,
  \clover\ will make polarimetric observations of approximately
  1000\,deg$^2$ of the sky in spectral bands centred on 97, 150 and
  225\,GHz.  The observations will be made with a two-mirror compact
  range antenna fed by profiled corrugated horns.  The telescope beam
  sizes for each band are 7.5, 5.5 and 5.5\,arcmin, respectively.  The
  polarisation of the sky will be measured with a rotating half-wave
  plate and stationary analyser, which will be an orthomode
  transducer.  The sky coverage combined with the angular resolution
  will allow us to measure the angular power spectra between $20 <
  \ell < 1000$.  Each frequency band will employ 192 single
  polarisation, photon noise limited TES bolometers cooled to 100\,mK.
  The background-limited sensitivity of these detector arrays will
  allow us to constrain the tensor-to-scalar ratio to $0.026$ at
  3$\sigma$, assuming any polarised foreground signals can be
  subtracted with minimal degradation to the 150\,GHz sensitivity.
  Systematic errors will be mitigated by modulating the polarisation
  of the sky signals with the rotating half-wave plate, fast azimuth
  scans and periodic telescope rotations about its boresight.  The
  three spectral bands will be divided into two separate but nearly
  identical instruments\,---\,one for 97\,GHz and another for 150 and
  225\,GHz.  The two instruments will be sited on identical three-axis
  mounts in the Atacama Desert in Chile near Pampa la Bola.
  Observations are expected to begin in late 2009.}

\section{Introduction}

The currently favoured cosmological model predicts that gravity waves
produced during a period of cosmological inflation should have
imprinted a faint primordial \mbox{``B-mode''} polarisation pattern in
the CMB\,\cite{hu03,seljak97}.  The amplitude of the gravity-wave
signal is related to the expansion rate, and hence energy
scale\,\cite{knox02}, during inflation and is therefore not yet
predicted by fundamental theory. By precisely characterising the
polarization of the CMB, it should be possible to put strong
constraints on the various theories of inflation. A discovery of the
primordial B-mode signal would be a significant breakthrough for both
astrophysics and particle physics, since we would then be probing
physics at the $10^{16}$\,GeV scale.

To date, the characterisation of the CMB temperature anisotropy has
given the prevailing $\Lambda$CDM cosmological model a strong
footing\,\cite{wmap3_spergel,wmap5_hinshaw}.  This footing was further
strengthened when the brighter ``\mbox{E-mode}'' polarisation signal,
which the prevailing model predicted, was recently
measured\,\cite{quad_1yr,capmap}.  The anticipated \mbox{B-mode}
signal has not yet been discovered because the current generation of
instrumentation is not sensitive enough to detect it.

In this paper, we describe \clover, which is a next-generation CMB
experiment that will have sufficient sensitivity to measure the T and
\mbox{E-mode} signals and, by searching for a cosmological
\mbox{B-mode} signal, measure or constrain the tensor-to-scalar
ratio\,\footnote{The tensor-to-scalar ration, $r$, is proportional to
  the energy scale of inflation to the fourth power\,\cite{knox02}.}
down to $r = 0.026$.  This is a factor of 20 lower than the current
upper limit from CMB measurements alone\,\cite{wmap5_dunkley}.
Therefore, the core astrophysical goals of \clover\ are the following:
to determine the energy scale of inflation, to improve constraints on
a collection of cosmological parameters, to measure a non-primordial
gravitaional lensing-induced \mbox{B-mode} signal, and to precisely
characterise any polarised Galactic signals such as synchrotron
radiation and polarised emission from dust.  An overview of
\clover\ is given in Section~\ref{sec:overview}, while instrument and
observation details are given in
Sections~\ref{sec:instrument}~\&~\ref{sec:observation}.

\begin{figure}[t]
\begin{center}
\epsfig{file=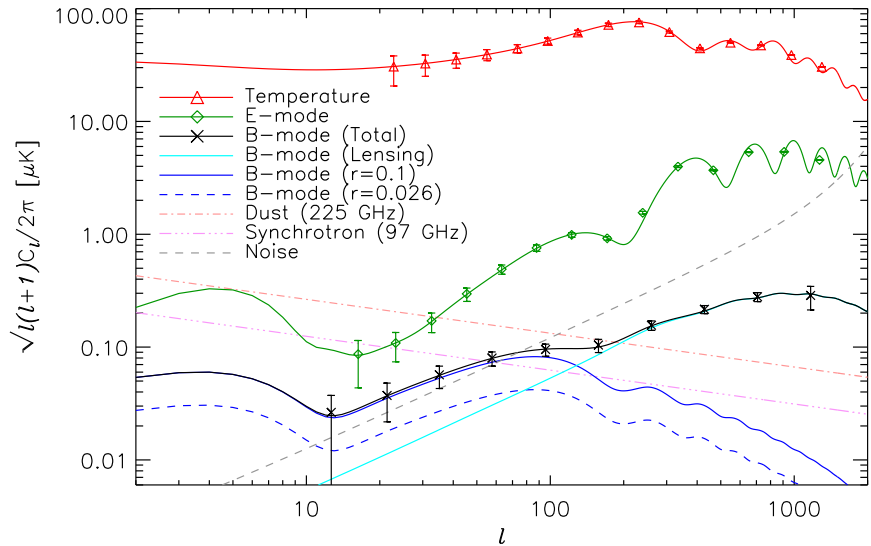,angle=90,width=120mm}
\caption{\footnotesize{Theoretical angular power spectra for the
    temperature, \mbox{E-mode} and \mbox{B-mode} signals assuming the
    $\Lambda$CDM cosmological model\,\protect\cite{wmap5_komatsu} and
    $r=0.1$.  The black, \mbox{B-mode} curve comprises a primordial
    inflationary gravity wave signal and a predicted non-primordial
    \mbox{B-mode} foreground signal produced by the gravitational
    lensing of partially polarised CMB radiation as it passed through
    large-scale structures.  The expected performance of \clover\ is
    overplotted as binned points, with error bars taking into account
    the atmospheric and instrumental noise for 150\,GHz and the sample
    variance for the \clover\ survey area.  For comparison, we plot
    the power spectra of the forecasted foreground signals in the
    \clover\ observation regions {\it before} cleaning, based on
    polarisation observations and models of unpolarised
    emission\,\protect\cite{wmap3_page,giardino02,sfd99}.  These
    foreground signals include polarised thermal emission from aligned
    dust grains and polarised synchrotron emission.  The dust and
    synchrotron signals are plotted at 225 and 97\,GHz, respectively,
    to show the worst case scenario.  Both signals should be smaller
    at 150\,GHz.}}

\label{fig:ps}
\end{center}
\end{figure}

\section{The \clover\ Experiment}
\label{sec:overview}

During two years of operation, \clover\ will make polarimetric
observations of approximately 1000 deg$^2$ of the sky from Pampa la
Bola in the Atacama Desert, Chile, which is within the ALMA science
preserve.  This location was selected because the millimetre-wave
transmittance of the atmosphere at the site is among the best in the
world for ground-based observatories, primarily due to the high
altitude of 4.9\,km.  The polarimetric observations will be made in
spectral bands centred on 97, 150 and 225\,GHz with two independent
instruments, one for 97\,GHz and another for 150 and 225\,GHz.  The
three frequency bands, primarily set by atmospheric transmission
windows, are well matched to the 2.7\,K CMB blackbody spectrum and are
expected to provide sufficient leverage for the spectral removal of
the anticipated astrophysical foreground signals.  The baseline design
of the experiment calls for the use of 576 superconducting transition
edge sensors (TES), cooled to 100\,mK with a pulse tube cooler, a He-7
refrigerator and a dilution refrigerator\,\cite{teleberg08}.  The
large number of detector outputs will be multiplexed with a
time-domain cryogenic readout.  The receivers containing the detectors
will be mounted at the focal planes of low cross-polarisation,
off-axis compact range antenna reflecting telescopes. The 97\,GHz
telescope will provide 7.5\,arcmin beams on the sky, while the
150/225\,GHz telescope will provide 5.5\,arcmin beams.  The expected
instrument NET $\simeq 18\,\mu \mbox{K} \sqrt{\mathrm{sec}}$ at
150\,GHz gives a predicted map sensitivity of around 1.7\,$\mu$K per
5.5\,arcmin resolution element for the $Q$ and $U$ Stokes parameters
after a two-year observing campaign.  The beam sizes are well matched
to the primordial \mbox{B-mode} signal, which dominates the
non-primordial \mbox{B-mode} signal below $\ell \simeq 100$ for $r
\gtrsim 0.01$, and will allow very good sensitivity to the
non-primordial \mbox{B-mode} signal from gravitational lensing (see
Figure~\ref{fig:ps}).  To mitigate the effect of 1/$f$ noise from
detector drifts and atmospheric fluctuations, to reject systematic
errors and to achieve the best noise performance, \clover\ will use
rapid polarisation modulation.  This modulation is produced by a
rotating achromatic half-wave plate and fixed orthomode transducers,
which are the polarisation analysers.  Because the \mbox{B-mode}
signal is faint, the instrument design outlined above has been advised
by detailed cosmological simulations to ensure our results should not
be contaminated by spurious instrumental signals\,\cite{odea_2007}.
The expected performance of \clover\ is shown in Figure~\ref{fig:ps},
and a model of one of the \clover\ instruments is shown in
Figure~\ref{fig:clover}.

\section{Instrument Description}
\label{sec:instrument}

\begin{figure}[p]
  \centering
  \epsfig{file=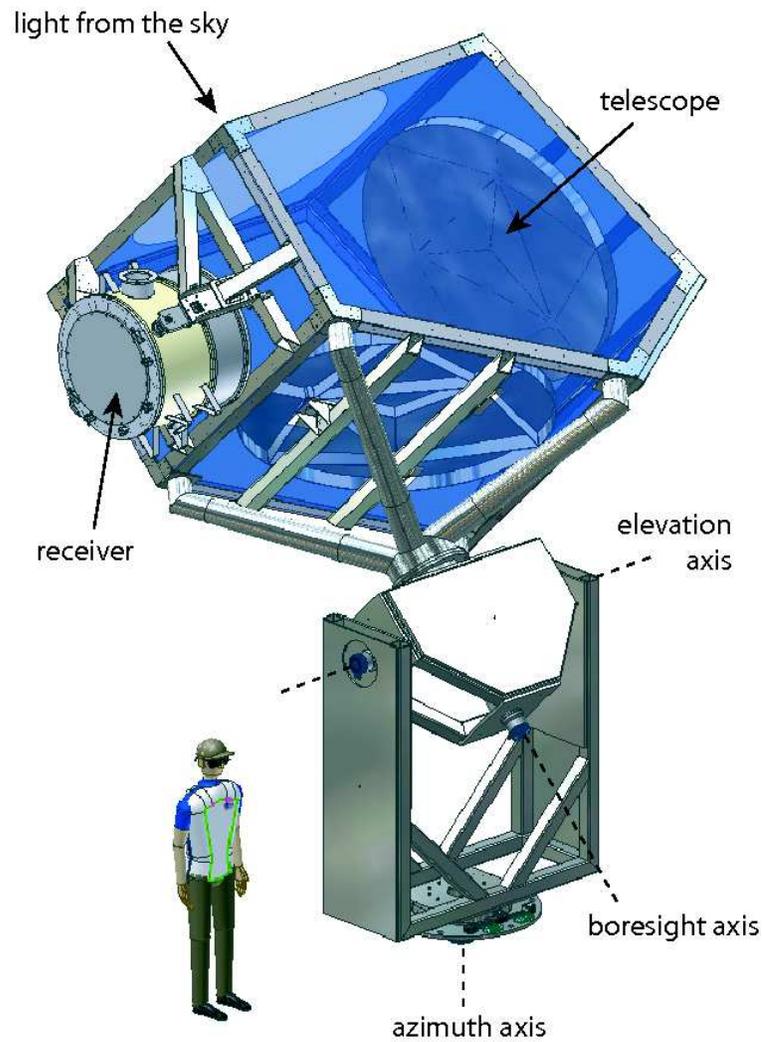,width=100.mm}
  \caption{\footnotesize{A model of one of the two
      \clover\ instruments. The three-axis mount also allows the
      rotation of the entire optical assembly around the telescope
      boresight.  The mirrors are held inside a co-moving baffle
      (translucent in the figure for clarity).  The co-moving baffle
      is lined with absorber to reduce the effect of side lobes.  A
      counter balance, which is not shown here, will be mounted on the
      boresight axis on the opposite side of the elevation axis from
      the telescope.  All of the instrument hardware shown here is
      either built or under construction.}}
  \label{fig:clover}
\end{figure}

Both the \clover\ instruments will use a compact range antenna (CRA),
which is composed of a parabolic primary mirror and hyperbolic
secondary mirror.  Both mirrors are off axis.  This optical design
gives excellent cross-polarisation performance and low aberrations
across a large, flat focal plane\,\cite{johnson_2008}.  For all focal
plane elements, the Strehl ratio is greater than 0.95, and the cross
polarisation\,---\,including the cross polarisation of the
horn\,---\,is better than $-38$\,dB.  The projected diameters of the
primary mirrors are 1.8 and 1.5\,m for the 97 and 150/225\,GHz
instruments respectively.  The telescope mirrors will be surrounded by
a co-moving baffle lined with millimetre-wave absorber, which will
prevent signals in the far side lobes of the telescope from being
modulated as the telescope scans.

Each array element in the focal plane comprises a profiled corrugated
horn, an orthomode transducer (OMT) and two (TES) detectors.  The TES
detectors are photon noise limited Mo/Cu superconducting bolometers
with transition temperatures around 190\,mK (430\,mK for 225\,GHz) and
time constants around
0.5\,ms\,\cite{audley2008_spie,audley2008_stt}. In the 97\,GHz focal
plane the OMT is composed of an electroformed turnstile junction with
a circular waveguide input and two rectangular waveguide
outputs\,\cite{pisano_2007}. Each output waveguide terminates in a
microstrip-coupled TES via a finline transition\,\cite{north_2008}.
At 150 and 225\,GHz each horn couples to a planar OMT composed of four
rectangular probes in a cylindrical waveguide\,\cite{grimes07_omt}.
Outputs from opposing pairs of probes are combined in a microstrip
circuit, which terminates in the TES; this design combines the OMT and
detectors on a single chip.  The focal plane elements are grouped into
eight-element modules, and each module is read out with a SQUID
multiplexer\,\cite{nist_ubc,reintsema03,korte03}.

A single achromatic half-wave plate (AHWP) will be mounted
approximately 100\,mm in front of each focal
plane\,\cite{pisano_2006,savini_2006}.  The AHWP is composed of three
A-cut sapphire discs at 97\,GHz and five discs at 150/225\,GHz.  The
diameter of each disc is approximately 300\,mm, and the disc
thicknesses are 4.65\,mm and 2.43\,mm at 97 and 150/225\,GHz
respectively.  The 97\,GHz AWHP central disc has a birefringent axis
oriented at 59\deg\ relative to the outer discs, while the second to
fifth crystals in the 150/225\,GHz stack are rotated by 29\deg,
95\deg, 29\deg and 0\deg\ respectively with respect to the first
crystal in the stack.  To improve transmission and to minimise
spurious polarisation from differential reflection, the AHWP will have
a three-layer, broadband anti-reflection coating on the external faces
of the stack.

The cryostat and optics are mounted to a common optical assembly,
which sits on top of a three-axis mount as shown in
Figure~\ref{fig:clover}.  The third axis, rotation around the
boresight, is necessary for identifying and suppressing systematic
errors from instrumental polarisation and for increasing
cross-linking.

\section{Observation Strategy}
\label{sec:observation}

\begin{figure}[t]
\begin{center}
\epsfig{file=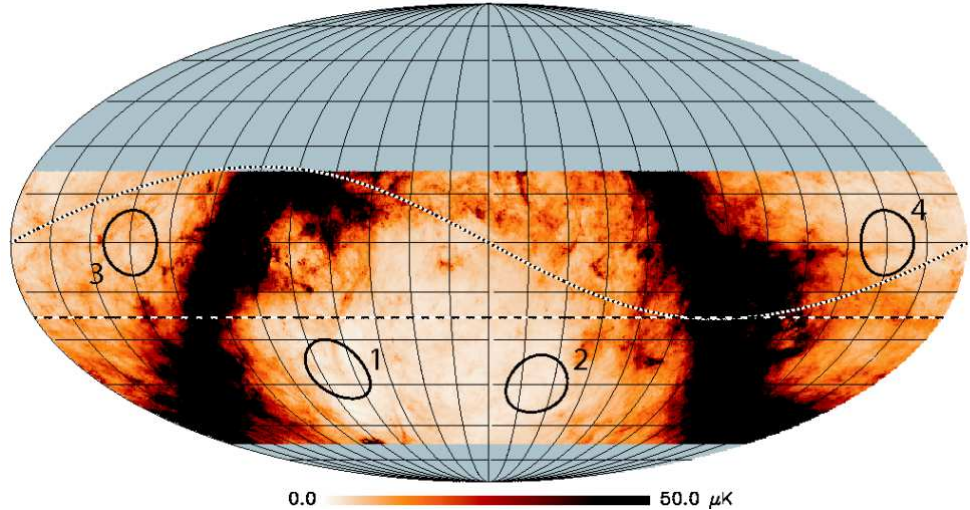,width=130mm}
\caption{\footnotesize{The four observation regions selected for
    \clover, centred on 0430$-$40, 2230$-$45, 0900$+$00 and 1400$+$00,
    plotted in equatorial coordinates over a model of the magnitude of
    the polarised Galactic emission at 225\,GHz.  The field locations
    were chosen based on models of the Galactic
    emission\,\protect\cite{giardino02,sfd99}, and are optimally
    distributed in RA.  These fields are intended to overlap nicely
    with the survey areas of other \mbox{B-mode} experiments.  Also
    plotted are the zenith declination (dashed) from Pampa la Bola
    (22\deg\,28$^\prime$\,S, 67\deg\,42$^\prime$\,W) and the ecliptic
    plane (dotted).  Grey areas never rise above 45\deg\ from the
    \clover\ site, which means they are not useful or accessible.}}
\label{fig:observations}
\end{center}
\end{figure}

\clover\ will observe a total of 1000\,deg$^2$ of the sky.  This
coverage is divided into the four regions shown in
Figure~\ref{fig:observations}, each roughly 20\deg\ in diameter.  Two
of the fields are in the southern sky and two lie along the equator.
To control the contribution from the atmosphere each telescope will
scan at approximately 0.5\,deg/sec in azimuth at a constant elevation.
Every few hours, the elevation angle of the telescope will be
re-pointed to allow for field tracking.  During all observations, the
AHWP will be rotated continuously at approximately 5\,Hz and the
telescope will be rotated around its boresight axis periodically.
Observations at 97\,GHz are expected to begin in late 2009, with
150/225\,GHz observations expected from mid-2010.

\section*{Acknowledgement}

\clover\ is funded by the Science and Technology Facilities Council.
CEN acknowledges an STFC studentship.  BRJ acknowledges an STFC
postdoctoral fellowship, and an NSF IRFP fellowship.  Some of the
results in this paper have been derived using the
HEALPix\,\cite{healpix} package.

\section*{References}
\bibliographystyle{moriond}
\begin{flushleft}
\begin{footnotesize}
\bibliography{abbrev_cen,moriond08}
\end{footnotesize}
\end{flushleft}
\end{document}